\documentstyle[prl,aps,epsf,floats]{revtex}
\begin{document}
\input{epsf}
\twocolumn[
\hsize\textwidth\columnwidth\hsize\csname@twocolumnfalse\endcsname
\draft 
]
\noindent{\bf Marinari et al. reply:} 

In a comment \cite{COMMENT} to our paper \cite{NOSTRO} H. Bokil,
A. Bray, B. Drossel and M. Moore claimed that we have reached wrong
conclusions.  We show here why their claims are not correct,
especially when compared to the analysis of reference \cite{LORO}.

As Bokil et al. correctly point out in \cite{NOSTRO} we based our 
analysis on the parameter

\begin{equation}
G=\frac{\overline{(\chi_{SG}^J)^2}-\bigl (\overline{\chi_{SG}^J}\bigr)^2}
{\overline{V^2\langle(q-\langle q\rangle)^4\rangle}-
\bigl(\overline{\chi_{SG}^J}\bigr)^2}\; .
\label{eqG}
\end{equation}
In the high-temperature phase $G$ goes to zero like $V^{-1}$, where
$V$ is the volume, while it should converge to $\frac{1}{3}$ in a low
temperature replica broken phase \cite{RIGOROUS}.  So this new
parameter plays the role of the Binder parameter or kurtosis in
ordered systems.  Obviously, in using $G$, one has to be sure that
the denominator is non-zero.  In one of the cases that we discussed
in \cite{NOSTRO} (finite dimensional Ising spin glasses with two spin
interaction) there is no problem since it is well known in the
literature that the denominator is non zero \cite{OLD}.

In our paper, instead of presenting the data for $G$, we could have
shown the data for another relevant parameter ($A$) defined as
follows:

\begin{equation}
A=\frac{\overline{(\chi_{SG}^J)^2}-\bigl(\overline{\chi_{SG}^J}\bigr)^2}
{\bigl(\overline{\chi_{SG}^J}\bigr )^2}\; .
\label{eqA}
\end{equation}
It is clear that $A$ does give the same kind of information carried by
$G$, i.e.  it signals the onset of a non self-averaging
susceptibility, and does not have the problem of a potentially zero
denominator that worries Bokil et al..

\begin{figure}
\centerline{\epsfxsize=8cm\epsffile{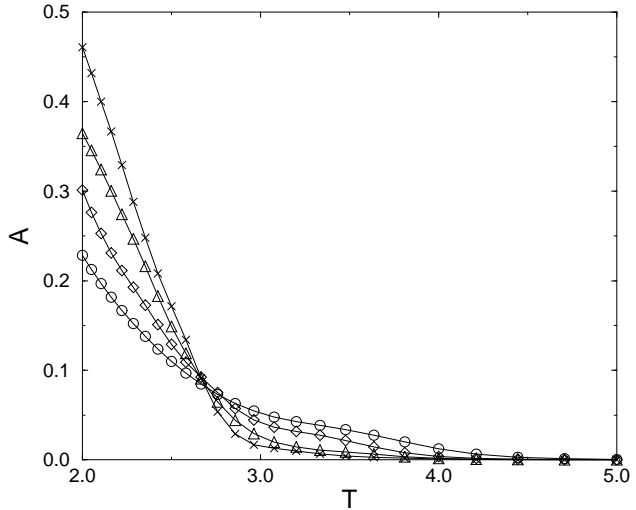}}
\caption{$A$ for the $3$-spin model. $L=3,4,5,6$ correspond to
circles,diamonds,triangles and crosses \label{F-F}}
\end{figure}

In figure (\ref{F-F}) we show as an example the results obtained for the
four dimensional $3$-spin model (the model of figure $4$ in
reference \cite{NOSTRO}).  The results of figure (\ref{F-F}) show
unambiguously that the different curves for $A$, corresponding to
different lattice sizes, cross at $T=T_c=2.62$, and are non-zero below
$T_c$: in this region $A$ increases with $L$.  The parameter $A$ is
a second good indicator for the transition.  If $A$ is not zero, as
happens in this case, the numerator and the denominator of equation
(\ref{eqG}) are both finite in the limit $L\to \infty$.

Although in a complete analysis one should study both parameters, for
lack of space it was impossible to present our data for both $A$ and
$G$.  In our paper we chose to discuss only the parameter $G$,
since the prediction about its asymptotic value in the low temperature
phase (one third) is useful to make the numerical analysis clear, and
because the result that the denominator is non zero is so well
established (at least in the most important case we were considering).
 
Now a few words about why we believe that the comment by Bokil et al.  is
confusing.  The reason for which Bokil et al.  find a non zero $G$ in a
situation where replica symmetry is non broken has nothing to do with a
small denominator (it is well known that the MK approximation does not
describe well spin glasses, since already in mean field it implies a
trivial droplet structure~\cite{GARDNER}, failing to describe the rich
structure of the SK model).

In three dimensional spin glasses in the MK approximation at $T=0.7$,
for reasonable values of $L$, the function $P_{L}(q)$ looks indeed like
a replica broken probability distribution and does not depend on $L$.
Only for extremely large values of $L$ one would see that in the MK
approximation replica symmetry is not broken: in the MK approximation
finite size corrections go to zero extremely slowly.  In this
situation it is reasonable that any estimator based on the behavior
of $P_{L}(q)$ will give the same misleading answer for reasonable
values of $L$.  This is what happens for the parameter $G$ and what
likely also happens with the parameter $A$.

The test of Bokil et al.  only tells us that in the crossover region
near the critical temperature where the behavior of the system (in
the MK approximation) is dominated by the critical point, the
parameter $G$ takes a value similar to the value at the critical
temperature.  They left unanswered the most interesting question,
which is the asymptotic behavior of $G$ in a model (like the MK
approximation) where replica symmetry is not broken.

There are two separate issues: (a) the existence of strong finite size
effects beyond the MK approximation; (b) the necessity of checking
that the denominator of eq.  (\ref{eqG}) is non-zero.  Issue (a) will
be carefully discussed in \cite{MPRZ} and issue (b) is easily solved
complementing $G$ with $A$.  Mixing the two issues together, as was done
in \cite{COMMENT}, leads to confusion.

\vspace{0.5truecm}
\bigskip

\noindent E. Marinari, C. Naitza, G. Parisi, M. Picco, 
F. Ritort and F. Zuliani\\ Barcelona, Cagliari, Paris and Roma\\
20 November 1998\\
PACS: 75.50.Lk, 05.50.+q, 64.60.Cn

\end{document}